\documentclass{sig-alternate}
\pdfoutput=1
\usepackage{enumitem}
\usepackage{setspace}
\usepackage{color}
\usepackage{url}
\usepackage{graphicx}
\usepackage{pdfpages}
\begin{document}


\title{A Methodology for Oracle Selection of Monitors and Knobs for Configuring an HPC System running a Flood Management Application}
\vspace{-7cm}
\numberofauthors{7}
\author{
\alignauthor
Panagiota Nikolaou \\
       \affaddr{University of Cyprus}
\alignauthor
Yiannakis Sazeides\\
       \affaddr{University of Cyprus}
\alignauthor Antoni Portero\\
 \affaddr{IT4Innovations, National Supercomputing Center}
\and
\alignauthor 
Radim Vavrik\\
     \affaddr{IT4Innovations, National Supercomputing Center}
\alignauthor Vit Vondrak\\
     \affaddr{IT4Innovations, National Supercomputing Center}
}

\maketitle

\begin{spacing}{0.865}
\begin{abstract}

This paper defines a methodology for the oracle selection of the monitors and knobs to use to configure an HPC system running a scientific application while satisfying the application's requirements and not violating any system constraints. This methodology relies on a heuristic correlation analysis between requirements, monitors and knobs to determine the minimum subset of monitors to observe and knobs to explore to determine the optimal system configuration for the HPC application. 


At the end of this analysis, we reduce an 11-dimensional space to a 3-dimensional space for monitors and a 6- dimensional space to a 3-dimensional space for knobs. 
This reduction shows the potential and highlights the need for a realistic methodology to help identify such minimum set of monitors and knobs.

%
%
%
%
\end{abstract}


%
%
%
%


\section{Introduction}

High Performance Computing (HPC) infrastructures run a diverse set of applications with different Quality-of-Service (QoS) requirements~\cite{brewer:2001}. These requirements come in various forms, such as operational power, performance, energy, cost and availability. Naturally, HPC systems need to be configured in a way to satisfy those application requirements~\cite{iordache2015}. To configure a system, all the different system hardware and software knobs are set at specific settings, with many remaining at a default setting, and then, while the application runs, various monitors are observed to determine whether the various requirements are satisfied. Unsurprisingly, the configuration space is extremely large and such configuration search efforts are in practice adhoc and non-optimal.
\begin{spacing}{0.95}
One way to reduce the configuration search dimensionality and complexity, it is to reduce the requirements, monitors and knobs that need to be satisfied, observed and explored, respectively. Reduction of a problem's dimensionality is not a new problem for computing system analysis~\cite{hoste:2007,hoste2006,hoste2006performance,annavaram2004}. Such reduction, typically, relies on some form of statistical correlation, for example, principal component analysis~\cite{jolliffe2002,abdi2010}.
\end{spacing}
In this work we explore the potential benefits, from a yet to be determined, realistic methodology that can identify the minimum set of monitors and knobs to use for configuring an HPC system that runs a specific application while satisfying the application requirements. To accomplish this, we use a correlation methodology that relies on data obtained from a detail exploration of a configuration space. Even though, a detail exploration is what a practical search methodology should avoid, the analysis in this paper, is useful to assess the potential benefits that can come from a future methodology that reduces the search space. \\
Specifically, for this investigation we use Floreon+ application, a flood prediction and management application, running on an HPC platform.
The analysis considers data obtained using eleven system monitors when exploring many settings for six knobs. This analysis is performed for individual and combinations of the following requirements: performance, power, availability, energy and cost. The results reveal that the configuration space can be reduced considerably. Additionally, the results show non-obvious correlations between requirements, monitors and knobs. This motivates the need for further research to determine a practical  configuration space reduction methodology for HPC systems.

The rest of the paper is organized as follows: Section~\ref{sec:Background} covers the background related to the Floreon+~\cite{portero2014system} application and HPC organization.  Section~\ref{sec:evaluation} describes the characterization and experimental framework and correlation analysis, and, Section~\ref{sec:results} presents and discusses experimental results. This paper concludes in Section~\ref{sec:conclusions}.



\begin{figure*}[h]
\centering
\begin{minipage}[b]{.42\textwidth}
\includegraphics[width=1\textwidth]{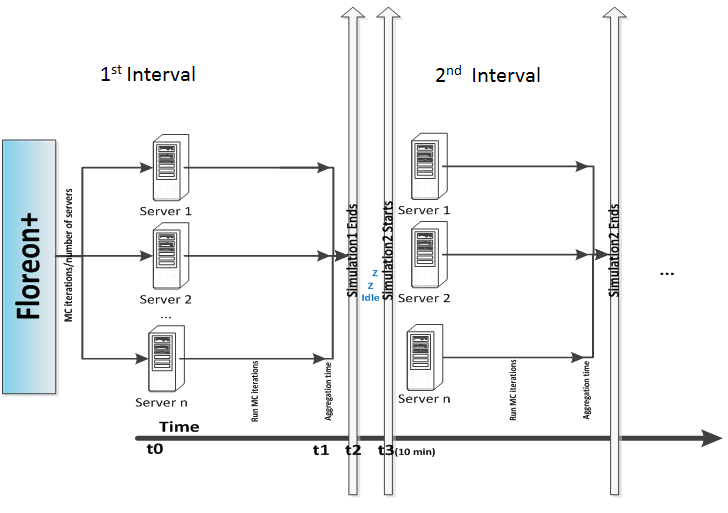}
\vspace{-0.7cm}
\caption{Floreon+ normal operation}
\label{fig:floreon_normal}
\end{minipage}
\begin{minipage}[b]{.5\textwidth}
\centering
\includegraphics[width=1\textwidth]{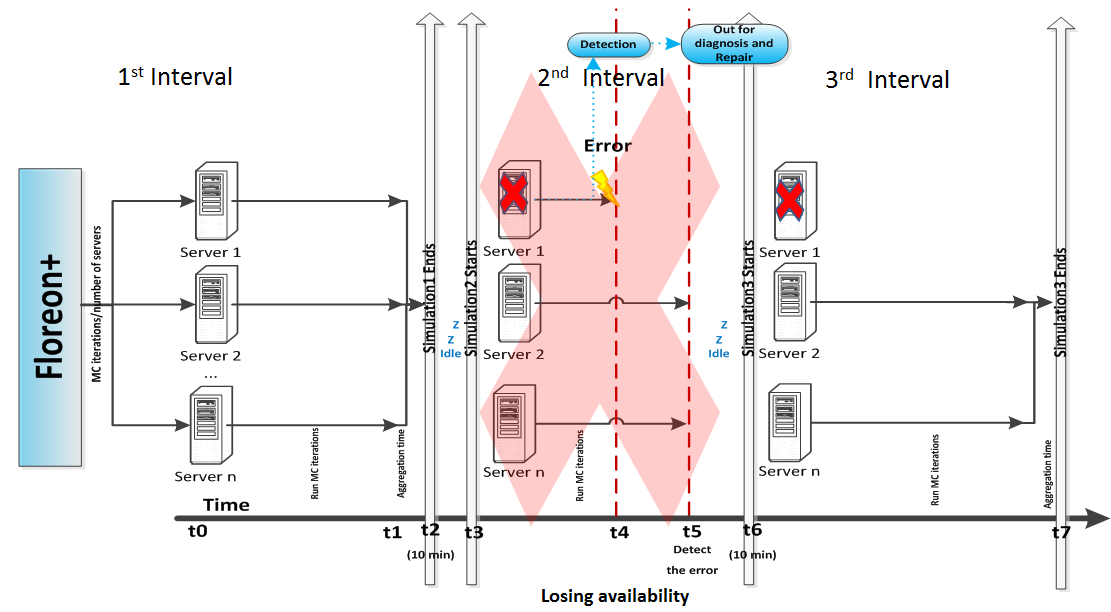}
\vspace{-0.7cm}
 \caption{Floreon+ operation with faults }
\label{fig:floreon_faulty3} 
\end{minipage}
\end{figure*}


\section{Background}\label{sec:Background}

\subsection{HPC Application (Floreon+)}
In this work, we use Floreon+ (FLOod REcognition On the Net)~\cite{portero2014system}, an HPC application with high QoS requirements. Floreon+ is an online system for monitoring, modeling, prediction and support disaster flood management~\cite{kuchar2015using}.  The system focuses on acquiring and analyzing relevant data in near-real time. The data are used to provide short-term flood prediction by running hydrologic simulations.

The main processes of Floreon+ application are organized as follows: 
 
\begin{enumerate}
\item Get information about actual river and reservoir situation.\item Rainfall-runoff (RR) modeling: simulation of surface runoff.\item Hydrodynamic (HD) modeling: flood lake simulations, flood maps, simulations of water elevation and water velocity, a real-time hydrological model for flood prediction, water quality analysis, etc.
\item Erosion modeling: simulation of water erosion.
\item Collection and archiving of flood data that can be used to estimate the magnitude of a flood based on historical evidence.
\end{enumerate}  
In this work we are investigating the uncertainty of the Rainfall-runoff (RR) modeling which is the most computationally intensive part~\cite{kuchar2015using,portero2014system}. 

The application framework for the uncertainty of RR model provides an environment for running multiple simulations every repetition, when new data arrives on an HPC system. The uncertainty contains information about how accurate is the solution that RR model provides. RR model is a dynamic mathematical model, which transforms rainfall to flow at the catchment outlet. 

The uncertainty is computed as Monte Carlo samples. The Monte Carlo method gives a straightforward way of massive parallelism by increasing the number of random values working concurrently to obtain numerical results. Previous experiments~\cite{kuchar2015using} exhibit a good scalability of the Monte Carlo method in an HPC cluster with 64 nodes of 16 cores each. Consequently, the proposed methodology in this paper may be appropriated in any other HPC framework where Monte Carlo method is employed.


Figure~\ref{fig:floreon_normal} shows the normal operation of Floreon+. As Figure~\ref{fig:floreon_normal} shows, a batch of Monte Carlo iterations is running in a number of nodes (Server 1- Server n) in such a way that application's QoS requirements are satisfied. Each interval indicates the execution of a different simulation. For example, the 1\textsuperscript{st} interval refers to the 1\textsuperscript{st} simulation. After the execution of all the Monte Carlo iterations, the results send to a master server for processing. The total simulation time includes the execution time of Monte Carlo iterations and the time needed to process the results. When a simulation ends, the servers remain idle for a set of period. The duration of this period is determined by the availability of the new batch of data. Under normal operation (fault-free) the simulation always finishes within the time constraint. There are some cases where a fault on a component can delay the execution  of the simulation, as shown in Figure~\ref{fig:floreon_faulty3}. These cases can be categorized in the following:
\begin{enumerate}
\item Delay the execution of the simulation but still the simulation finishes within the time constraint. The availability of the system does not decrease.
\item Delay the execution and violate the timing constraint with the same number of servers.  Thus the results of this simulation are useless and the availability of the system decreases.
\item Delay the execution and violate the timing constraint with less servers. 
In this case the faulty server needs to be taken offline until it is repaired or replaced. In this case the results are lost and the availability decreases. 
\end{enumerate}  

Figure~\ref{fig:floreon_faulty3} illustrates the last case where the server needs to be taken offline until it is repaired or replaced. As shown in the figure, the number of servers in the 3\textsuperscript{rd} interval is decreased and the job is assigned to the remaining servers until the faulty server is repaired or replaced. 

Next we present the various requirements for this application.

\subsubsection{Application Requirements}
Floreon+ has two running operation modes, the standard operation mode and the emergency operation mode. Both have different requirements. Standard operation mode is the default operation of the system. In this operation the weather is favorable and the flood warning level is below of the critical threshold. 

On the other hand, on the emergency operation mode the water in the rivers rises due to continuous rain or free-flowing streams that are created due to heavy rainfall on small areas. During this operation mode much more accurate and frequent simulation computations are needed and the results should be provided as soon as possible.   
In this work we focus on the emergency operation which has tighter timing requirements.  

\subsubsection{Availability and Quality of Service requirements}
The reliability and availability target of Floreon+ running on emergency operation is accomplished through a combination of hardware/software mechanisms and policies. This also aims at satisfying the QoS requirements even in the presence of errors and offline servers. These mechanisms typically rely on hardware and software \textbf{monitors} and \textbf{knobs}. 

In general, when a server fails and if its repair time is expected to be long, the system software migrates the failed job to another server. The failure of the server is detected by a hardware or software monitor. The migration is possible because for QoS and availability reasons. HPC systems are over-provisioned with spares for dealing with errors and offline servers. Server over-provisioning is determined by the availability of a system. The less available system, the more servers needed~\cite{Nikolaou:2015}

Because of its significance, emergency operation requires responsiveness in 10 minutes for each simulation.  
Also it must provide high levels of availability, two nines (0.99), which may require over-provisioning with extra servers to deal with various hardware failures.

Floreon+ and other offline services can run together (collocated) to improve utilization~\cite{Mars:2011,Yang:2013}. Specifically, when Floreon+ satisfies the QoS requirements without using all the available cores in a server, the remaining cores can run other services. This must be done without affecting the QoS of Floreon+ and violating its requirements. Since we are going to explore the emergency operation, Floreon+ is running in isolation, utilizing all the available resources without any other service concurrently runs on the same server. 

\subsubsection{Accuracy on Monte Carlo Iterations}

It is of utmost importance that the results are as precise as they can. The precision of the simulated results is based on the number of Monte Carlo iterations~\cite{portero2015flood}. It has been shown in~\cite{portero2015flood} that the number of iterations has to be in the order of $10^{4}$ to $10^{5}$ to obtain a satisfying precision. In this work, we assume 20000 Monte Carlo iterations that has to be computed before the deadline (i.e. ten minutes, since new input data arrives from weather stations) as the baseline configuration. 

\begin{table}[t]
\center
\caption{Floreon+ requirements}
\begin{tabular}{|l|l|}\hline
Performance &simulation $\leq$ 10 minutes\\\hline
Accuracy &$\geq$ $10^{4}$ MC iterations \\\hline
Availability   &$\geq$0.99 \\\hline
Power & $\leq$81 Watts\\\hline
Energy & $\leq$48600 Joules\\
\hline
\end{tabular}
\label{tab:applic_requirements}
\end{table}

Table~\ref{tab:applic_requirements} summarizes the specific values for the requirements of the Floreon+ application. 


\subsection{Available Monitors and Knobs in HPC Systems}
A large number of monitors and knobs exist in HPC systems. Monitors enable the observation of physical, micro-architectural, and operating system phenomena that can assess the status of a system as well as the progress towards completing a task. On the other hand, knobs enable the proactive or reactive control of various phenomena. 

Monitors and knobs in real systems can be categorized into the following categories depending on what metric they influence: performance, power, temperature and reliability.

Table~\ref{tab:Mon_Knobs} shows the requirements, monitors and knobs that are going to be explored in this work. This subset is by no means comprehensive and future work should consider a larger set.

\begin{table}[t]
\center
\caption{Application requirements, Monitors and Knobs }
\begin{tabular}{ |c|c|c| }\hline
Requirements &Monitors &Knobs \\\hline
Performance &Execution Time &DVFS\\
Power &IPC &SMT\\
Energy &DRAM Power &DRAM Protection\\
Availability &CPU Power &Turbo Mode\\
Cost &Peak Power &Prefetchers\\
		&CPU Temperature &Redundancy\\
		&MPKI &\\
		&Server MTBF &\\
		&System MTBF &\\
		&Capex expenses & \\
		&Opex expenses & \\\hline
\end{tabular}

\label{tab:Mon_Knobs}
\end{table}

\begin{table}[t]
\center
\caption{Server Configuration}
\begin{tabular}{|l|l|}\hline
Number of CPUs  &2\\
CPU             &Intel Xeon E5-2640 v3 \\
Number of cores per CPU  &8\\
Number of threads per core  &2\\\hline
Channels per CPU        &4\\
DIMMs/channel &2\\
DIMM capacity &16GB\\\hline
\end{tabular}
\label{tab:server_conf}
\end{table}

Monitors that are going to be investigated are: execution time, Instructions per Cycle (IPC),  Misses per kilo Instructions (MPKI), DRAM, CPU and peak power and CPU temperature. Also, Mean Time Between Failures (MTBF) per server and for the whole system is used through analytical models and Failures In Time (FIT) rates~\cite{Sridharan:2013}. Finally, capital (CAPEX) and operational (OPEX) expenses are going to be estimated based on publicly available info, e.g. list prices, and runtime measurements. CAPEX expenses include infrastructure, server and networking equipment costs, whereas OPEX expenses include power and maintenance costs.

Table~\ref{tab:Mon_Knobs}, also shows the different knobs that we are going to experiment with.  As the table shows, we use  Simultaneous Multi-Threading (SMT)~\cite{Tullsen:1995}, Dynamic Voltage and Frequency Scaling (DVFS)~\cite{Weiser:1994}, data prefetchers~\cite{tendler2002} and Intel's Turbo Mode~\cite{turbo_boost}.
Also, this work provides results with and without redundant cores. Redundant cores are used to improve the reliability of the system by migrating the running thread of a faulty core to a spare~\cite{huang2010}. For the redundancy scenario it is assumed that half of the cores remain idle to provide higher availability. On the other hand, in the scenario without redundancy all the available physical resources are utilized. Furthermore, this study explores the implication of using two different DRAM Protection Techniques (No Protection or ChipkillDC).

DRAM is protected from errors by using extra devices per DIMM to store
Error Correction Code (ECC) bits. Modern processors usually support Chipkill with 16 ECC bits to protect 128 data bits that are interleaved across two DIMMs placed in two
channels~\cite{Amd2010,HP:new,Amd2014}. This is referred as ChipkillDC or Lockstep where it can correct all the errors in a single device and detects all the
errors in two devices~\cite{Amd2010}. Chipkill can waste bandwidth, hurt
performance and increase energy consumption~\cite{Ahn2009,Jian2013,Udipi2012}. On the other hand No Protection does not provide any protection on DRAM.


%
%

\begin{table}[t]
\center
\caption{Values of knobs}
\begin{tabular}{ |l|l|l|}\hline
Knobs &Value\\\hline
DVFS  &1.2, 1.7, 2.2, 2.6 (GHz) \\\hline
SMT &Disable, Enable \\\hline
DRAM Protection &No Protection, ChikillDC \\\hline
Turbo Mode &Disable, Enable\\\hline
Prefetchers &Disable, Enable\\\hline
Redundancy &Disable, Enable \\\hline
\end{tabular}
\label{tab:Knobs}
\end{table}



\section{Experimental Framework and \\Correlation Analysis}\label{sec:evaluation}

\subsection{Experimental Framework}\label{sec:exp}
For evaluation, we use an HPC cluster with dual socket Intel Xeon E5-2640 v3 system configuration, as shown in Table~\ref{tab:server_conf}. 
We run each experiment 5 times, and each time we observed execution time, instructions per cycle (IPC), misses per kilo instructions (MPKI), CPU, DRAM power and CPU temperature. The results presented are calculated by removing minimum and maximum values and calculating the average.

Prefetchers and DRAM protection techniques changed by accessing BIOS, through a BIOS
Serial Command Console interface (CLI)~\cite{Guide:2014}.

Our evaluation used \textit{Floreon+}, an HPC application with a dataset of 44KB (Tiny dataset). This is a representative dataset size for the application purposes because it uses five days observations to provide predictions for the next two days. All the power numbers are collected using the Likwid-powermeter~\cite{likwid} which allows monitoring the power consumption of CPU and DRAM at any given time. The results are used to calculate total power and peak power numbers.

To track CPU temperature, we use lm-sensors~\cite{lm-sensors}. 
To estimate Server MTBF and System MTBF monitoring values, as well as, availability values we use different analytical models based on binomial probabilities. 

Also, to estimate CAPEX, OPEX expenses as well as total cost we use COST-ET and AMPRA tools proposed in~\cite{Hardy:tco,Nikolaou:2015}. 

For each experiment, an initial population of 2 server modules is used. The number f servers can change depending on availability and performance requirements.

For a baseline configuration, we select the one that is currently used to run this application and includes the following values for each parameter: SMT:OFF, Frequency: 2.6 GHz, DRAM Protection: No Protection, Turbo Mode: Enable, Redundancy: 0 (No), Prefetchers: ON. 

The data used for correlation analysis are obtained by exploring the 128 combinations of knobs presented in Table~\ref{tab:Knobs}. For each configuration combination the eleven monitor values are recorded. These eleven values are then used to determine the values of five metrics each corresponding to a specific requirement. 

The values of the requirements for a given configuration are obtained as follows: performance from execution time, power from cpu and dram power, energy product of execution time and power, availability from server MTBF, server MTTR and number of servers, and cost using monitors for power, performance, availability using the COST-ET tool~\cite{Hardy:tco}. The total data set used for this analysis is, therefore, a 128 x 11 x 5 data matrix, where 128 are the combinations of knobs, 11 are the monitors values and 5 are the requirements.

In this Section we describe the methodology used for correlation analysis to reduce the number of requirements, monitors and knobs used to configure an HPC system. The data that drive this analysis are obtained as discussed in Section~\ref{sec:exp}.
The value obtained for each requirement for a given experiment, are normalized to give each requirement equal weight. The normalization is done by computing the mean and standard deviation for each requirement and then by subtracting the mean and dividing the standard deviation by each value of requirements as in~\cite{hoste:2007}. A configuration is considered to be successful if it satisfies each individual application requirement. To rank successful configurations we determine an overall score for a configuration 
by multiplying each normalized requirement value with an equal weight. In this case, since we have 5 requirements, we multiply each value with 0.2.
The correlation analysis is done using the R statistical language~\cite{team2013r}.

\subsubsection{Heuristic Correlation Analysis}

This analysis explores the correlation between requirements, monitors and knobs. The methodology used is as follows:
\begin{enumerate}
\item For a given requirement, we compute the correlation coefficient (using Pearson correlation analysis) with all the other requirements. For each pair (x$_i$,x$_j$) of requirements i and j, where i$\neq$j, that exhibit significant correlation coefficient (above a 90\% threshold \footnote{\label{note1}The specific thresholds are picked on empirical analysis not shown here}), we check which of the two requirements can be removed. The requirement that shows smaller correlation coefficient with all the other requirements is removed from the list. This process is iterated over all remaining requirements. The same process is repeated for the monitors.
\item For the remaining requirements and monitors, we compute the correlation coefficient between each requirement and all remaining monitors and select the monitor with the highest correlation. 
\item For all the remaining monitors and all the available knobs we compute the correlation coefficient between them and knobs that have a correlation coefficient above a 40\% threshold \textsuperscript{\ref{note1}} are kept.  
 \end{enumerate}

This correlation analysis aims to reduce the number of configurations that need to be explored to determine the configuration that provides the highest satisfaction of the requirements according to the ranking in Section~\ref{sec:exp}. Specifically, the analysis returns a subset of the monitors and knobs. All possible configuration combinations are then evaluated for the selected knobs. For the knobs that are not selected we used the baseline configuration values (we have confirmed that it does not really matter which value you use for them). Afterwards, the selected configurations are sorted according to the selected monitor(s)  value(s) (if there is more than one monitor, equal weighting is used to combine them). The top ranked configuration using the selected knobs and monitors is then compared with the configuration that considers all. Their difference is measured as the maximum negative \% difference for any of the requirements (if it is 0 for all it means it matches the best possible configuration).

\begin{table}[t]
\center
\caption{Correlation between Monitors and Requirements}
\begin{tabular}{ |l|l|l|}\hline
Requirements &Monitors\\\hline
Performance  &Execution time \\
Availability &Capex cost \\
Power &CPU power \\
Cost &Capex cost\\
All  &Execution time, CPU power, Capex cost \\\hline
\end{tabular}
\label{tab:correlation_mon_req}
\end{table}

\section{Results}
\label{sec:results}

 \subsection{Results of the Heuristic Correlation Analysis}
Our analysis reveals that Energy is strictly correlated with performance and thus we removed Energy from the list of requirements and the presented results.

Table~\ref{tab:correlation_mon_req} shows the results of correlation analysis between requirements and monitors.

As we can see from the Table, Performance requirement can be monitored by Execution Time. Availability can be, monitored by Capex expenses. This correlation is not obvious  because Capex cost has an indirect relation with Availability. 
The correlation of Capex expenses and availability comes from the extra servers that are needed for over-provisioning. The more servers  needed for over-provisioning, the more capex expenses are incurred. Moreover, the more servers are needed for over-provisioning, the less the System MTBF due to the higher probability of errors. So, this leads to a smaller availability for this system. 
As the Table, also, shows, CPU power is the appropriate metric for the total power. This happens because Floreon+ is not a memory intensive application so it has a negligible impact on DRAM power. For the total cost and overall combination of requirements, capex is the correlated monitoring value.

\begin{figure}[t]
 \begin{minipage}[b]{.5\textwidth}
 \centering
\includegraphics[width=1\textwidth]{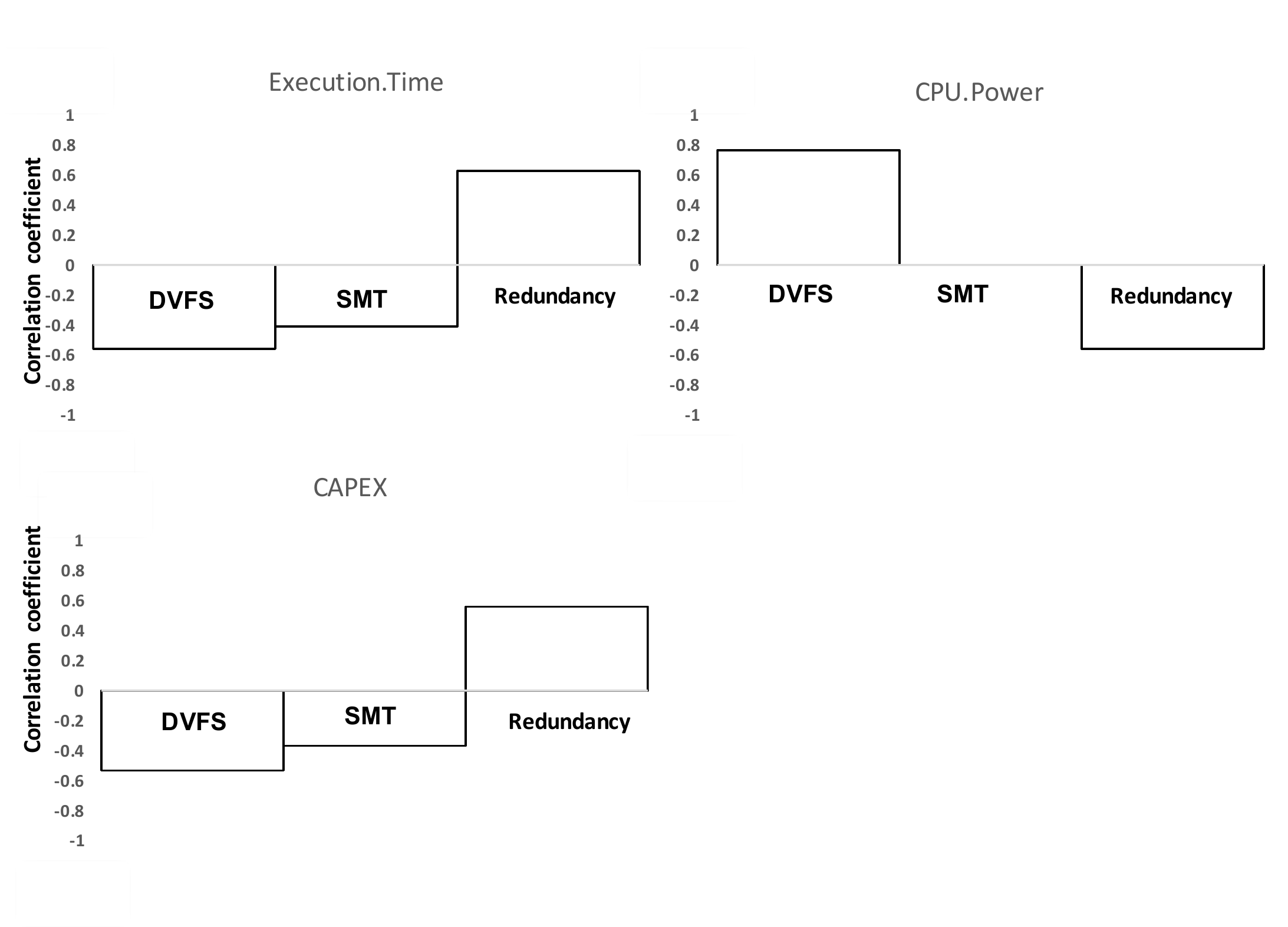}

 \caption{{Correlation analysis presenting the correlation coefficient between Monitors and Knobs}}
 \label{fig:corr_mon_knobs}
 \end{minipage}
  \end{figure}

Finally, Figure~\ref{fig:corr_mon_knobs} shows, the correlation between knobs and the monitors.
X-axis presents the remaining list of knobs for each monitor, whereas the y-axis presents the correlation coefficient between knobs and monitors. The correlation coefficient ranges from -1 to +1. A value closer to +1 means that this knob has almost a linear relation with the monitor. A value closer to -1 means that this knob has an inverse relation with the monitor. A value closer to 0 means that there is no correlation between the knob and monitor.\\ 
As we can see from the Figure, DVFS, SMT and Redundancy are the selected knobs. On the other hand, DRAM Protection, Turbo Mode and Data prefetching can be reduced from the search space. This is due to the low cache and low memory pressure of the Floreon+ application (L3MPKI: 0.0034,  L2MPKI: 0.013).

 \subsection{Validation of Heuristic Methodology}

\begin{figure}[t]
 \begin{minipage}[b]{.5\textwidth}
 \centering
\includegraphics[width=1\textwidth]{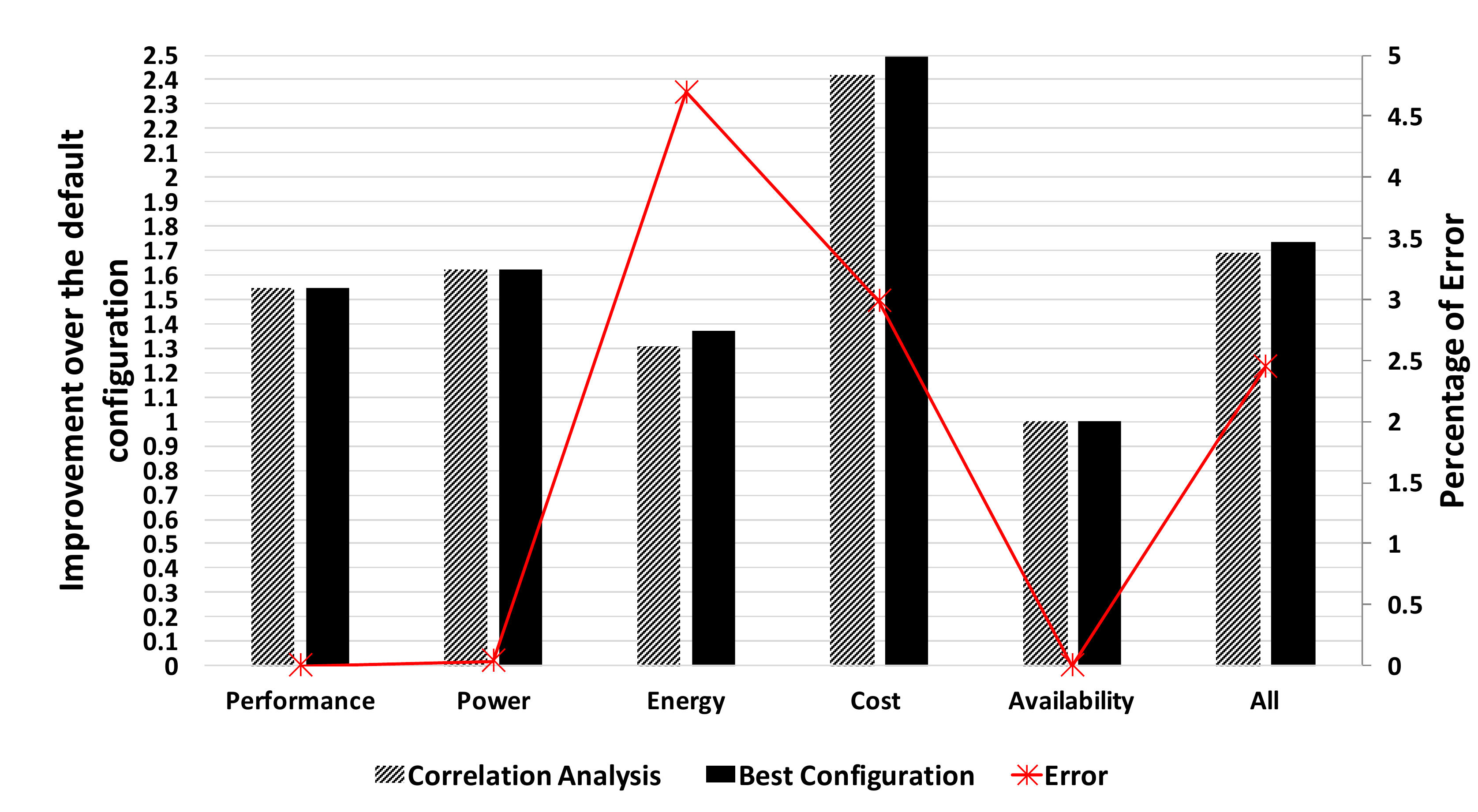}

 \caption{{Improvement between the default configuration, correlation analysis predicted configuration and the best configuration for different requirements}}
 \label{fig:prediction}
 \end{minipage}
  \end{figure}

Figure~\ref{fig:prediction} validates the heuristic analysis by evaluating the best configuration for each requirement using the subset of monitors and knobs
 revealed from the correlation analysis.

In this graph, we show the normalized with the default configuration, improvement for each individual requirement and all requirements between the best configuration using the selected monitors and knobs and the best configuration using all monitors and knobs.  As can be seen from Figure~\ref{fig:prediction}, all the requirements can be improved by changing the system configuration. For example, performance can be improved by 1.5x and cost by 2.4x related to the default configuration. Also, the combination of all the requirements can be improved by 1.7x. Moreover, the Figure shows that the configurations based on the correlation analysis are very close to the results with the best configuration. We measure the error between the two configurations and we found a maximum value of 5\%.

 \section{Conclusions and Future work}\label{sec:conclusions}
 This paper describes an oracle methodology that investigates combinations of different knobs and the effects they have on the various monitored values. This aims to explore the potential benefits, from a yet to be determined, realistic methodology that can reduce the number of monitors and knobs that configure an HPC system that runs a specific application while satisfying the application requirements.
To this end, this analysis reduces an eleven space of system monitors to three and a six system space knobs to three. 
 
Findings of this work motivates future research. One direction is to find a realistic methodology probably based on correlation analysis that will be generic for many applications. Another direction is to minimize the data sample used.

 \section{Acknowledgment}\label{sec:acknowledgment}
The research leading to this paper is supported by the ``Harpa, Project No 612069"  of the European Commission 7th RTD Framework Program, the ``Uniserver, Project No 688540" of the European Community's H2020 Program,  the Czech Republic Ministry of Education, Youth and Sports from the Large Infrastructures for Research, Experimental Development and Innovations project, IT4Innovations National Supercomputing Center -- LM2015070 and the University of Cyprus.
\end{spacing}

\begin{spacing}{0.83}
\bibliographystyle{abbrv}

\bibliography{references}  
\end{spacing}

\end{document}